\documentclass[preprint2]{aastex}

\slugcomment{Not to appear in Nonlearned J., 45.}

\shorttitle{Elemental Abundances of Polar Jets}
\shortauthors{Lee et al.}

\begin{document}

\title{Photospheric Abundances of Polar Jets on the Sun Observed by Hinode}

\author{Kyoung-Sun Lee}
\affil{Hinode team, ISAS/JAXA, 3-1-1 Yoshinodai, Chuo-ku, Sagamihara, Kanagawa 252-5210, Japan}
\email{lksun@solar.isas.jaxa.jp}

\author{David H. Brooks \altaffilmark{1}}
\affil{College of Science, George Mason University, 4400 University Drive, Fairfax, VA 22030, USA}
\altaffiltext{1}{Current address: Hinode Team, ISAS/JAXA, 3-1-1 Yoshinodai, Chuo-ku, Sagamihara, Kanagawa 252-5210, Japan}

\and
\author{Shinsuke Imada}
\affil{Solar-Terrestrial Environment Laboratory (STEL), Nagoya University, Furo-cho, Chikusa-ku, Nagoya 466-8550, Japan}


\begin{abstract}
  Many jets are detected at X-ray wavelengths in the Sun's polar regions, and the ejected plasma along the jets has been suggested to contribute mass to the fast solar wind. From in-situ measurements in the magnetosphere, it has been found that the fast solar wind has photospheric abundances while the slow solar wind has coronal abundances. Therefore, we investigated the abundances of polar jets to determine whether they are the same as that of the fast solar wind. For this study, we selected 22 jets in the polar region observed by Hinode/EIS (EUV Imaging Spectrometer) and XRT (X-Ray Telescope) simultaneously on 2007 November 1-3. We calculated the First Ionization Potential (FIP) bias factor from the ratio of the intensity between high (S) and low (Si, Fe) FIP elements using the EIS spectra. The values of the FIP bias factors for the polar jets are around 0.7-1.9, and 75$\%$ of the values are in the range of 0.7-1.5, which indicates that they have photospheric abundances similar to the fast solar wind. The results are consistent with the reconnection jet model where photospheric plasma emerges and is rapidly ejected into the fast wind.
 \end{abstract}


 \section{Introduction}
The solar wind is characterized by two streams depending on their speed, fast and slow, which show different plasma properties. The fast solar wind has a lower proton density and a cooler temperature, determined from the freezing-in temperature, while the slow solar wind has a higher proton density and a hotter temperature. Their plasma compositions are also different: the composition of the fast solar wind is close to that of the solar photosphere while the composition of the slow solar wind is similar to the corona \citep{geiss95, vonsteiger00}. Because of their different plasma properties, it is thought that the formation and acceleration mechanisms of the slow and fast solar wind are different, and investigations of their source regions are important. 
From Ulysses observations, it has been found that the fast solar wind emanates from the polar regions while the slow solar wind comes mostly from equatorial regions \citep{mccomas98}. Based on these results, it is suggested that polar coronal holes, polar jets, polar plumes and inter-plumes are sources of the fast solar wind, and equatorial streamers and the boundaries of active regions might be slow wind source regions. However, it still remains unclear and needs to be investigated further.

To search for fast and slow solar wind source regions, the difference in elemental composition is a useful parameter. The differences between photospheric and coronal abundances can be related to the `FIP (first ionization potential) effect'. Elements with a FIP below and above 10 eV are termed `low FIP' and `high FIP' elements, respectively. In the slow solar wind, the low FIP elements are enhanced by factors of 3-4 relative to their photospheric abundances, while the ratio of the low FIP elemental abundance to photospheric is approximately equal for the fast solar wind. With remote spectroscopic observation of the solar atmosphere, such as from Skylab and SOHO (Solar and Heliospheric Observatory), there have been attempts to obtain the abundances of discrete solar structures and candidates for solar wind source regions \citep{feldman98, feldman05,  feldman00, feldman03, raymond97, wilhelm98, widing86, widing01, young99}. For example, \citet{feldman98} determined the FIP bias factors of a north polar coronal hole and a quiet equatorial region using SUMER (Solar Ultraviolet Measurements of Emitted Radiation: \citet{wilhelm95}), and showed that the results are in good agreement with the in-situ measurements in the fast and slow solar winds. \citet{raymond97} measured the elemental composition in streamers using the UVCS (Ultraviolet Coronagraph Spectrometer: \citet{kohl95}) and found that the abundances of the edge of the streamers are similar to the abundances in the slow solar wind. For the specific source region for the fast solar wind, \citet{wilhelm98} and \citet{young99} measured the abundances of plumes and inter-plumes above the polar region using SUMER and CDS (Coronal Diagnostic Spectrometer: \citet{harrison95}) and showed that plumes have only a small FIP effect over the photospheric value, even though the low-FIP elements were enhanced compared to the inter-plume.


After the launch of Hinode \citep{kosugi07}, determination of the spatially resolved abundances of specific regions became possible using high sensitivity and high resolution EUV spectroscopy \citep{feldman09, brooks11, brooks12, baker13}. For example, using the Hinode/EUV Imaging Spectrometer (EIS: \citet{culhane07}), \citet{brooks11} determined the FIP bias factor for the outflowing active region boundary which was suggested as a possible solar wind source region by \citet{sakao07}, \citet{harra08}, and \citet{doschek08}. They found that the active region outflows show coronal abundances, which is consistent with the abundance of the slow solar wind. These methods can also be applied to specific features in the polar regions to provide further insights. 

Jets in polar coronal holes are one of the candidates to be a source of the fast solar wind along with coronal holes, plumes, and inter-plumes. With Yohkoh/Soft X-ray Telescope (SXT) and Hinode/X-ray Telescope (XRT: \citet{golub07}) observations, many x-ray jets are observed in the polar coronal holes. \citet{shibata92} suggested that the jets are caused by magnetic reconnection, and \citet{cirtain07} observed a fast velocity component ($\approx 800 \rm{km ~ s^{-1}}$) of the X-ray jet. From their observed speeds and masses, they suggested that jets can contribute to the solar wind and that reconnection can accelerate the plasma into the solar wind. Even though the abundances in the coronal holes, plumes, and inter-plumes were measured in previous studies \citep{feldman98, feldman05, wilhelm98, young99}, the abundances in polar jets have not been investigated yet.  
 In the present study, therefore, we investigate the elemental composition of polar jets using the Hinode/XRT and EIS and find that their abundances are similar to the abundances of the fast solar wind. 

\section{Observations and method}

 We investigated coronal jets seen in the north polar region of the Sun during 2007 November 1-3. The jets were observed by Hinode/XRT and EIS simultaneously. During the period, there are XRT images taken through the Al\_poly (November 1, 3)  and Al\_mesh (November 2) filter, which are sensitive to hotter plasma ($\approx$ 2 MK) and quiescent coronal emission. The XRT field of view is $512 \arcsec \times 512 \arcsec$ and its pixel resolution is about $1 \arcsec$. The time cadences of the XRT Al\_poly and Al\_mesh filter observations are about 50 seconds and 40 seconds, respectively. For the EIS observations, the polar region was observed using the study, `gdz$\_$plume1'. This study was designed for observing the coronal hole region using the 2$\arcsec$ slit with 4$\arcsec$ steps and 50s exposure time at each step. The field of view of the raster is $296 \arcsec \times 512 \arcsec$. This study has 23 spectral windows with a wide spectral range and the spectral lines we used are listed in Table 1.
 
 For the first step, we searched for jets in the XRT images based on their intensities and the shapes of the structures. Second, we coaligned the EIS scanning rasters with the XRT images which were observed within the EIS scan duration. We de-rotated the XRT images using the EIS observing time as a reference. Then, we obtained the offset values between the XRT Al\_poly filter images and the EIS \ion{Fe}{12} 195 $\rm{\AA}$ intensity images by cross correlation using the procedure `$align\_map.pro$' in the Solar Software (SSW) package. Finally, we checked the slit position during the raster, and selected jets which are caught by the EIS slit (Movie 1; Figure 1). 

Most of the jets have bright emission at the base, which corresponds to X-ray bright points (BPs), and collimated ejections are seen in the intensity images and Doppler velocity maps. However, some of the jets are very faint in the intensity images, even though they can be seen in the Doppler velocity maps with a clear signature of outflows along the jet structure. Faint jets like these can be categorized as "dark jets", the discovery of which has recently been reported by \citet{young15}. Therefore, we selected jets based on not only the intensity images, but also the EIS Doppler velocity maps. Figure 1 shows examples of two jets, one is on the solar disk and the other is on the limb. The first and second columns of Figure 1 show the XRT intensity image and \ion{Fe}{12} 195.12 $\rm{\AA}$ intensity image, respectively. Doppler velocity maps from EIS are shown in the third column. One can see that the jet has a clear Doppler velocity signature even in the case that the jet shows only weak intensity along the jet structure (3-Nov example).

We selected 22 jets observed by XRT and EIS and we obtained the intensities of the EIS spectral lines for the jets and nearby background regions by making masks for the bright point (`bp') at the base, and three parts of the jet regions which we label: `jet 1' (middle), `jet 2' (top), and `jet 3' (end of the jet). The solid and dotted contours in the EIS intensity images in Figure 1 correspond to the jet and background regions where we obtained the spatially averaged spectrum. Since the emission from the jets is weak one might suspect that it comes from the surrounding polar coronal holes. Therefore, we subtracted the background emission to isolate the jets. The forth columns in Figure 1 show the \ion{Si}{10} spectra of the jet (solid line) and background regions (dashed line) for the `jet 1' region of the jet on the disk (\#7) and the `bp' region of the jet on the limb (\#17). These figures show that the intensity is enhanced in the jet relative to the background. We subtract the background spectrum from the jet spectrum, then we fit the subtracted spectrum with single or double Gaussians. Figures 2, 3, and 4 exhibit examples of the fitted spectral line profiles of the subtracted spectrum for the jets in Figure 1. Due to the fall off of intensity and the resultant increase of the errors, we could only obtain the background-subtracted intensities for 22 `bp' regions at the bottom of the jets, 11 `jet 1' regions, and 5 `jet 2' and `jet 3' regions. For example, while we obtained enough signal in the 'jet 1' (middle) region and 'jet 2' (top) region of the jet \#7 (Figures 2 and 3), we were only able to obtain the subtracted intensity for the `bp' region for the dark jet \#17 due to the weak intensity along the jet (Figure 4). 
   
We calculated the FIP bias factor from the ratio of the intensity between high (S) and low (Si, Fe) FIP elements using the EIS spectral lines. \citet{feldman09} suggested several line pairs in the EIS spectral range as FIP bias indicators for the relative abundance determination. \citet{brooks11} developed the method using the spectral line pair of \ion{Si}{10} 258.37 $\rm{\AA}$ and \ion{S}{10} 264.23 $\rm{\AA}$, and we used their method to measure the FIP bias factors of the polar jets. 

First, we measured the differential emission measure (DEM) using the \ion{Fe}{8} - \ion{Fe}{15} line intensities, intensity errors, and the density measurements from the \ion{Fe}{13} 202.04 $\rm{\AA}$ to 203.82 $\rm{\AA}$ line ratio. 
For measuring the line intensities, we applied the revised radiometric calibration presented by \citet{delzanna13} because the EIS detector's response shows a degradation with time in the different wavelength channels after the launch of Hinode \citep{delzanna13, warren14}. For the error of intensities, we considered the 1 sigma error from the gaussian fit for the intensity measurement and a calibration error for the instrument ($\sim$23$\%$; \citet{lang06}). For the calculation of the DEM, we adopted the ionization fractions of \citet{bryans09} and the photospheric abundances of \citet{grevesse07} and used the Markov-chain Monte Carlo (MCMC) method in the PINTofALE spectroscopy package \citep{kashyap98} and the CHIANTI v7.1.3 atomic database \citep{dere97, landi12}. Figure 5 shows examples of the emission measure (EM) distribution from the DEM we calculated using the Fe lines. The blue and grey lines represent the best-fit solution from the MCMC method and its 1-sigma errors, respectively.  

Second, we scaled the DEM derived from the Fe lines using the \ion{Si}{10} 258.37 $\rm{\AA}$ intensity. 
We estimated the intensity of the \ion{S}{10} 264.23 $\rm{\AA}$ line using the scaled DEM and measured the ratio of the predicted intensity to the observed intensity, which represents the FIP bias factor. The detail of this method can be found in \citet{brooks11} and \citet{brooks15}. The sensitivity tests by \citet{brooks15} show that the method is robust, and is not significantly affected by calibration errors, differences in the fractionation level of Fe and Si, or uncertanties in the atomic data. In any case, in this study, the difference between the scaled DEM and the Fe DEM is less than 30\% for nearly all the examples we present (87\%).

Table 2 shows the measured FIP bias factors using this method. 
The estimated errors from the ratio of the predicted intensity of \ion{S}{10} using the scaled DEM from the low FIP elements (Fe, Si) and the observed intensity of the \ion{S}{10} are given. To estimate the error of the predicted intensity of \ion{S}{10}, we calculated the DEM error assuming the simulated DEMs are distributed uniformly within the upper and lower limits of the DEM calculated from the 90$\%$ confidence interval from the MCMC computation. For the error of the observed intensity, we measured the 1-sigma error of the intensity measurement. Then, we measured the error of the ratio between the predicted and observed intensities by combining their errors in quadrature. From the DEM analysis, we also obtained the peak EM temperatures of the jets for each region using a gaussian fit to the DEM. The results are shown in Figure 6. 

Note that polar jets are not steady structures: statistical studies show that they have velocities of $10-1000  \rm{km~s^{-1}}$ \citep{shimojo96, cirtain07, savcheva07}. For the DEM analysis, we assume ionization equilibrium, but, the plasma is highly dynamic, and there is a possibility that this assumption might be violated by the high speed plasma. 
To check whether the ionization equilibrium assumption is applicable or not, we calculated the time-dependent ionization fraction of Si and S using the average density of the jets ($10^{9} \rm{cm^{-3}}$) for a plasma temperature change from 1MK to 1.58MK using the method of \citet{imada11}. From the calculation, the plasma reaches ionization equilibrium in 100 seconds. The time difference between the first appearance of the jets in XRT and EIS is about 2 minutes or more for most jets (jet appearance timings in Table 2), so the plasma is likely to be in ionization equilibrium when the jets are observed by the EIS slit. Even for the other cases, the assumption is likely to be valid because the ratio of the ionization fraction of \ion{Si}{10} and \ion{S}{10} varies below the instrumental uncertainty (23 \%) most of the time. 



 \section{Results and Discussion}
 \subsection{Characteristics of the Jets: Density, Temperature, and Velocity}
 Figure 6 shows the characteristics of the jets we analyzed. The black, green, and red lines in Figure 6 indicate the `bp', `jet 1', and `jet 2 and jet 3' regions of the jets, respectively. The left and middle panels present the distributions of the density and temperature. The densities from the BPs of the jets are higher than those of the other parts of the jets. The average value of the density is about $ 9.3 \times 10^{8}  \rm{cm^{-3}}$. The peak EM temperatures of the jets are in the range $ 0.8-2.0 \times 10^{6} \rm{K} $, with an average value of about $ 1.4 \times 10^{6} \rm{K}$. The density and temperature values of the jets are similar to EIS measurements of polar jets reported in previous studies \citep{doschek10, young14}.

We also tried to estimate the propagating velocity ($ \approx 230 \rm{km~s^{-1}}$) using a combination of the projected velocity ($230 \rm{km~s^{-1}}$) and the Doppler velocity $15 \rm{km~s^{-1}}$) for a standard jet (\#17 in Table 2). The projected velocity is determined by a time-distance diagram along the jet axis using the XRT intensity images and the Doppler velocity derived from \ion{Fe}{12} $195.12 \rm{\AA}$. With the density and the propagating velocity of the jet, we roughly calculated the mass flux to compare with the solar wind mass flux. For this calculation, we used equation 1 from \citet{pucci13} and the jet occurrence rate from \citet{cirtain07} (240 per day) and \citet{savcheva07} (60 per day). We assume that the life time of the jet is about 10 min and the size of the jet is about 30\arcsec. 
The estimated flux is $ 1-4 \times 10^{7} \rm{cm^{-2} ~ s^{-1}} $ at the Earth distance, which is $2-20 \%$ of the solar wind mass flux at 1AU ($ 2-4 \times 10^{8} \rm{cm^{-2} ~ s^{-1}} $), which reported by previous studies \citep{feldman78, wang93, cranmer99}. The value is comparable to the mass flux ($ 4 \times 10^{7} \rm{cm^{-2} ~ s^{-1}} $) reported by \citet{cirtain07}\footnote{The value of the proton flux at the Earth given in the paper is actually $1 \times 10^{12} \rm{m^{-2}~s^{-1}}$ and \citet{young15} corrected the value to $4 \times 10^{11} \rm{cm^{-2} ~ s^{-1}} $. However, when we calculated the proton flux using the parameters in the paper, the proton flux is about $4 \times 10^{7} \rm{cm^{-2} ~ s^{-1}} $.}. We assume that all the jets are ejected to interplanetary space, however, this might not be the case \citep{paraschiv15}, and some material from the jets falls back to the solar surface \citep{culhane07b}. Therefore, the mass flux from jets in general may be less than our calculation. We also simply used the propagating velocity of one jet out of our sample, so, the mass flux still has a large uncertainty.

\subsection{Elemental Composition: FIP Bias Factor}
Our most significant new result is the measurement of the FIP bias factor. The right panel of Figure 6 shows the histogram of the FIP bias factor of the jets and BPs and the relationship with the jet regions. The FIP bias factors are in the range $0.7 - 1.9$, with an average value of about 1.2. The range of the FIP bias factors is quite wide, and we can see that some regions have higher values larger than 1.6. However, most of the FIP bias factors ($\approx 75 \%$) fall in the $ 0.7-1.5 $ range (0.3-2.2 considering the errors)
which is close to photospheric. When we measure the abundances for the BPs at the bottom of the jets and along the jets separately, the FIP factors of the BPs are in the range of 1.1-1.8 with an average value of 1.24 and those along the jets are within the range of 0.7-1.9 with an average value of 1.22. The FIP bias factors for the ejected jets and BPs are similar. The FIP bias factors of each region are given in Table 2. 

Our abundance measurements for polar jets imply the following. First, the FIP bias factors show that  
the abundances of the polar jets and BPs related to those jets 
are photospheric.  
The photospheric abundances of the polar jets and their BPs are consistent with the typical elemental composition of the fast solar wind reported from in-situ measurements \citep{geiss95, vonsteiger00}, suggesting that they could indeed be a fast wind source. They are, however, transient events, while the fast solar wind is a relatively steady phenomena, and it is unclear how such transient contributions can be smoothed out. One possibility is that stream interactions provide the smoothing before any jet contribution reaches 1AU, as found for fine scale surface flux-tube variations in the MHD simulations of \citet{cranmer13}. To establish the detailed relationship between jets and the fast solar wind, we need further investigations considering the connection of magnetic field lines to the near Earth and the response of the in-situ measurements.

Second, previous studies of the temporal variation of the FIP bias factor in several active regions have shown that newly emerged active regions have photospheric abundances and then, around 1 or 2 days later, their abundances reach coronal values \citep{widing01, feldman03}. Moreover, older active regions develop a higher coronal FIP bias factor over time (though see \citet{baker15}). In most of the cases of the polar jets and their BPs we analyzed, the flux emerges, then likely reconnects quickly within several tens of minutes so that the abundances remain photospheric. The measurements suggest that the whole process from emergence to reconnection of the jets occurs on a time scale shorter than is needed to change the abundances. 

Third, the result is also consistent with the reconnection jet model. \citet{shibata92} and \citet{yokoyama95, yokoyama96} suggested that emerging flux reconnects with overlying ambient field lines and the plasma is rapidly ejected along the newly reconnected field lines. Recently, \citet{pariat09} performed a numerical simulation for polar jets and showed that massive jets can be ejected at the Alfv\'enic velocity. Even if the emerging plasma is ejected quickly but at sub-Alfv\'enic velocity, the plasma of the jets will have photospheric abundances, which is in agreement with our measurements.

Furthermore, when we take a closer look at the higher values ($1.6 - 1.9$) of the FIP factors in our result, they come from several jets which occurred at the same location repeatedly. For example, jets \#1 to \#4 in Table 2 occurred at the same bright point which lasted for more than 3 hours. The FIP bias factors from those jets show slightly higher values compared to other jets, which may be an indication that the abundances are beginning to change, though the uncertainties are probably too large to be sure. The results imply that the jet plasma is not normally confined long enough to change the abundances from photospheric to coronal, but is also new indirect evidence that the confinement time is critical for the process, as expected by e.g. \citet{laming04}.

\section{Summary}
We have determined the characteristics of polar jets using Hinode/XRT and EIS observations. For this, we analyzed 22 jets observed on 2007 Nov 1-3. We measured the density, peak EM temperature, and propagating velocity of the jets using EIS and XRT. In particular, we measured the relative abundances of the polar jets. The FIP bias factors derived in the present study are in the range 0.7-1.9 and 75\% of the FIP factors are less than 1.6, which is close to photospheric. These values are consistent with the abundances of the fast solar wind suggesting that the jets can, in principle, contribute to the fast solar wind though no direct link has been established. 

\acknowledgments
Hinode is a Japanese mission developed and launched by ISAS/JAXA,
with NAOJ as domestic partner and NASA and STFC (UK) as
international partners. It is operated by these agencies in
cooperation with ESA and the NSC (Norway). 
CHIANTI is a collaborative project involving the following Universities: Cambridge (UK),
George Mason and Michigan (USA). 
This work was partially supported by the JSPS Core-to-Core Program 22001.
The work of DHB was performed under contract with the Naval Research Laboratory 
and was funded by the NASA Hinode program.  

\clearpage
\begin{deluxetable}{c c}
\tablecaption{List of the Spectral Lines Used  in the Present Study \label{tbl-1}} \tablewidth{0pt} \tablehead{
   \colhead{Line ID ($\rm{\AA}$)} & \colhead{$\log T_{max}$ (K)}}
    \startdata
        \ion{Fe}{8} 185.21 & 5.7  \nl 
        \ion{Fe}{10} 184.54 & 6.1 \nl
        \ion{Fe}{11} 188.22 & 6.2 \nl
        \ion{Fe}{11} 188.29 & 6.2 \nl
        \ion{Fe}{12} 186.88 & 6.2 \nl
        \ion{Fe}{12} 195.12 & 6.2 \nl
        \ion{Fe}{13} 202.04 $\ast$ & 6.3 \nl
        \ion{Fe}{13} 203.83 $\ast$ & 6.3 \nl
        \ion{Fe}{14} 264.79 & 6.3 \nl
        \ion{Fe}{14} 270.52 & 6.3 \nl
        \ion{Fe}{15} 284.16 & 6.4 \nl
        \ion{Si}{10} 258.37 & 6.2 \nl
        \ion{S}{10} 264.23 & 6.2 \nl
    \enddata
    \tablecomments{$ $ The peak formation temperature of the spectral lines are from the Chianti version 7.0. Lines used for the density determination are marked with asterisks. }
\end{deluxetable}

\clearpage
\begin{deluxetable}{c c c c c c c c c c c c c c c c }
\rotate
\tabletypesize{\footnotesize}
\tablecaption{List of Polar Coronal Jets Analyzed in the Present Study and their FIP bias factors \label{tbl-2}} \tablewidth{0pt} 
\tablehead{
  \colhead{\#} & \colhead{Date} & \multicolumn{3}{c}{XRT} &  \colhead{} &\multicolumn{2}{c}{EIS} & \colhead{} &\multicolumn{2}{c}{${\rm{Position} ^{1}}$} & \colhead{} & \multicolumn{4}{c}{FIP bias factor}\\ 
    \cline{3-5}  \cline{7-8} \cline{13-16} \\
    \colhead{} & \colhead{} & \colhead{Filter} & \multicolumn{4}{c}{Time (UT)$^{2}$} & \colhead{Doppler} & \colhead{} & \multicolumn{2}{c}{(arcsec)} & \colhead{} & \colhead{bp} & \colhead{jet1} & \colhead{jet2} & \colhead{jet3} \\
    \cline{4-7} \\
    \colhead{} & \colhead{} & \colhead{} & \colhead{t$_{xrti}$} &  \colhead{t$_{xrtf}$} &  \colhead{} & \colhead{t$_{eis}$} & \colhead{} &\colhead{} & \colhead{x} & \colhead{y} & \colhead{} &  \colhead{} & \colhead{} & \colhead{} & \colhead{} }
    \startdata  
   1 	& 2007/11/01 & Al\_poly & 00:38:05 & cont. & & 00:40:31 & blue & &135 & 960 & &1.45$\pm 0.30$ & 1.19$\pm 0.24$ &  & \nl 
   2	&		        & Al\_poly & cont. & cont. &  & 01:45:30 & blue & &\multicolumn{2}{c}{$\#1$} && 1.70$\pm 0.47$ & 1.05$\pm 0.22$ &  & \nl
   3   &	                & Al\_poly & cont. & cont. &  &  02:49:37 & blue&&\multicolumn{2}{c}{$\#1$} & &1.54$\pm 0.42$ & 1.58$\pm 0.57$ & 0.71$\pm 0.62$   & \nl
   4   &		        & Al\_poly & cont. & cont. & & 03:56:20 &blue &&\multicolumn{2}{c}{$\#1$} && 1.52$\pm 0.57$ & 1.62$\pm 0.67$ &  & \nl
   5   & 		        & Al\_poly & 04:02:24 & 04:43:56 & &  04:08:27 & red & & 85    & 970 & &1.15$\pm 0.20$ & & & \nl 
   6   & 		        & Al\_poly & cont. & cont. &  & 04:58:44 & blue & &\multicolumn{2}{c}{$\#1$} & &1.77$\pm 0.46$ & 0.75$\pm 0.26$ & & \nl
   7   & 2007/11/02	& Al\_poly & 03:42:32 & 04:05:02 &   & 03:47:44 & blue & & 35 & 840 && 1.19$\pm 0.25$ & 1.10$\pm 0.39$ &  & \nl
   8   &	 		& Al\_mesh & 11:32:54 & cont. & & 12:03:02 & blue & &65  & 790 & &1.32$\pm 0.23$ & 		 & 	      & \nl
  9  & 			& Al\_mesh & 14:54:55 & 15:27:36 & & 15:11:04 &blue& &\multicolumn{2}{c}{$\#8$} & &1.10$\pm 0.18$ & 1.16$\pm 0.37$ & 1.70$\pm 0.45$ & 0.98$\pm 0.71$ \nl   
  10  & 			& Al\_mesh & 15:34:56 & cont. & & 15:42:13 &blue & &-45  & 820 & & 1.40$\pm 0.57$ & 1.91$\pm 1.98$ & 1.45$\pm 0.71$ & \nl
  11  & 			& Al\_mesh & cont. & cont. & &16:13:29 & blue & &\multicolumn{2}{c}{$\#8$} & &1.34$\pm 0.23$ &  &  & \nl
  12  & 			& Al\_mesh & cont. &18:34:14 & &16:44:37 &blue & &\multicolumn{2}{c}{$\#10$} & &1.35$\pm 0.54$ & & & \nl
  13 & 			& Al\_mesh & 18:17:33 & 18:37:34 & & 18:24:22 &blue & & 80 & 840 & &1.57$\pm 0.31$ &  &  & \nl
  14 & 			& Al\_mesh & cont. & cont.  & & 19:25:53 & blue &&\multicolumn{2}{c}{$\#8$} & &1.62$\pm 0.25$ & 1.37$\pm 0.29$ &  & \nl  
  15 &	 		& Al\_mesh & cont. & cont. & & 20:30:01 & blue &&\multicolumn{2}{c}{$\#8$} & &1.74$\pm 0.40$ & 1.47$\pm 0.30$ & 1.36$\pm 0.46$ & \nl
   16 & 2007/11/03	& Al\_poly & 12:33:03 & 12:50:21 & &12:32:11 & blue && 30     & 970 & &1.32$\pm 0.45$ &  &                 & \nl
   17 & 			& Al\_poly & 13:56:54 & 14:06:55 &&14:04:52 & blue &&-100  & 950  & &1.78$\pm 0.71$ & & & \nl
   18 & 			& Al\_poly & 14:35:17 & 14:58:36 & &15:07:15 & blue &&\multicolumn{2}{c}{$\#17$} & &1.41$\pm 0.47$ & & & \nl
   19 &			& Al\_poly & 14:34:37 & cont. && 15:15:54 &red& &-125 & 970 & &1.21$\pm 0.24$ & & & \nl
   20 &			& Al\_poly & 16:01:05 & cont. && 16:20:02 &red &&\multicolumn{2}{c}{$\#19$} && 1.10$\pm 0.19$ & & & \nl
   21 & 			& Al\_poly & 16:28:47 & 16:53:28 & &16:39:10 & blue && 90 & 810 & &1.26$\pm 0.33$ & 0.87$\pm 0.17$ &  & \nl 
   22 & 			& Al\_poly & 18:12:42 & 18:36:02 & &18:26:17 & blue & &\multicolumn{2}{c}{$\#21$} && 1.16$\pm 0.30$ &  & & \nl
     \enddata
    \tablecomments{$^{1}$ Several jets occur at the same position. In these cases we marked the position with the number of the jet that occured at the same location. }
    \tablecomments{$^{2}$ t$_{xrti}$, t$_{xrtf}$, and t$_{eis}$ are the timings of the appearance and disappearance of the jets in XRT images and slit timings in EIS scan rasters, respectively. Several jets are recurrent continuously at the same location. For those cases, we marked the timings as `cont.'. }
    \end{deluxetable}

\clearpage
\begin{figure*}
\epsscale{1.6}
\centering
\includegraphics[height=70mm]{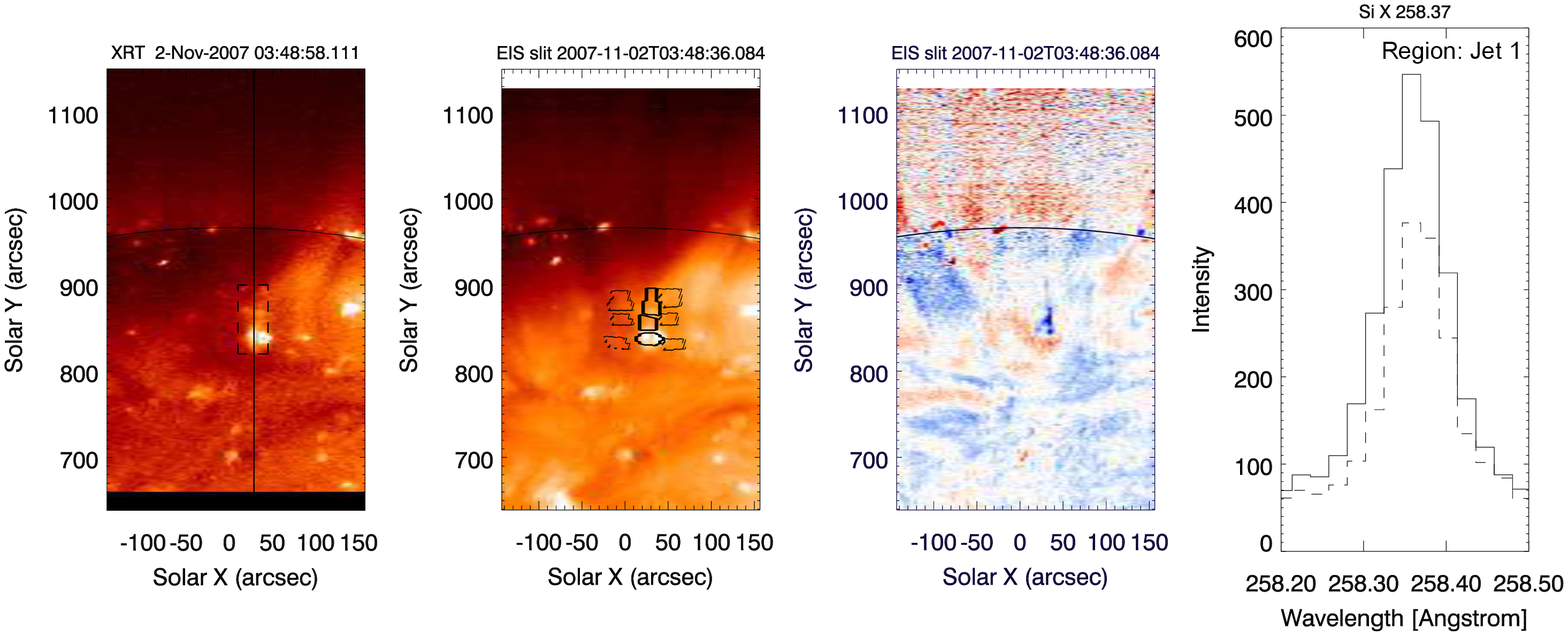}
\includegraphics[height=70mm]{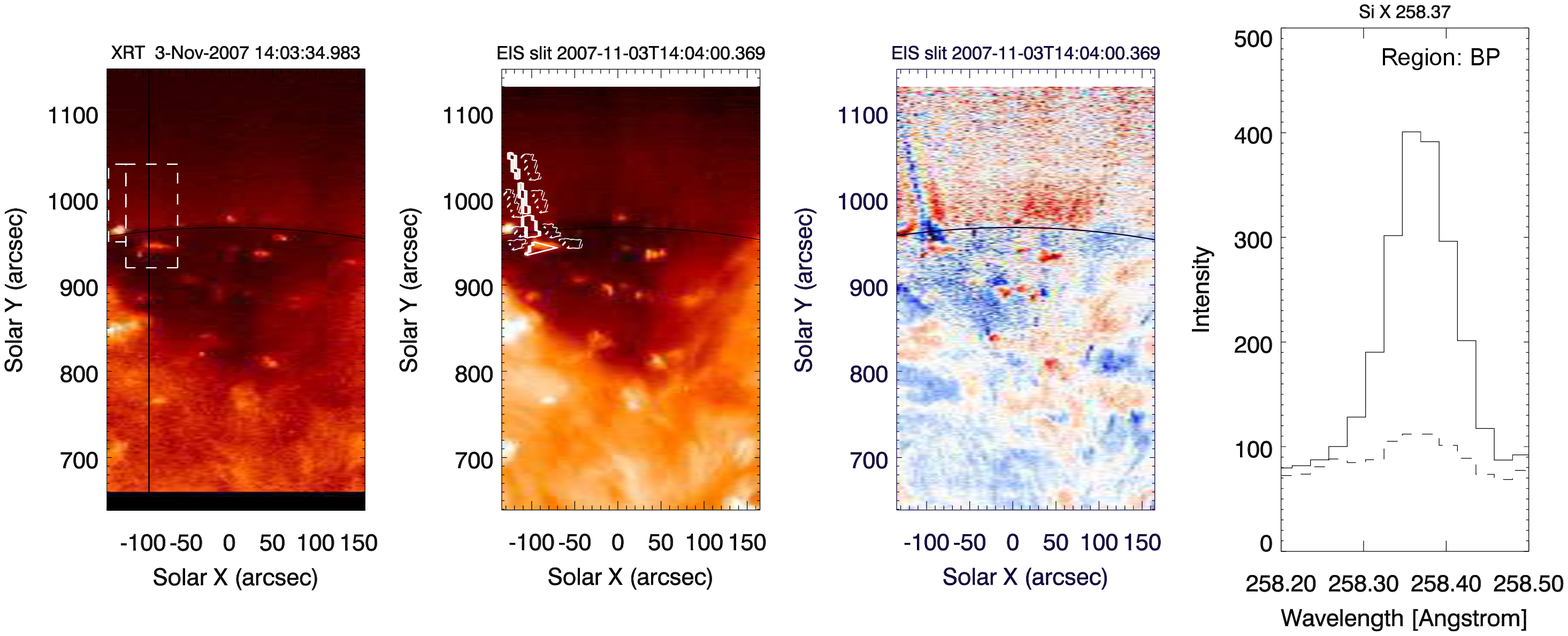}
\caption{Example of a polar X-ray jet observed by XRT and EIS (top: \#7, bottom: \#17 in Table 2). First column: XRT Al\_poly intensity images. The dashed box indicates the jet we analyzed. The vertical line is placed at the EIS slit location when the jet is observed. Second column: EIS \ion{Fe}{12} 195.12 $\rm{\AA}$ intensity images. Third column: EIS  \ion{Fe}{12} 195.12 $\rm{\AA}$ Doppler velocity images. The solid contours in the second panels are regions (`bp', `jet 1', `jet 2', and `jet 3') of the jet where we obtained the averaged spectrum for analysis. The dotted contours indicate the background regions for each jet. Fourth column: the \ion{Si}{10} spectra of the `bp' (for the jet \# 17) and `jet 1' (for the jet \# 7) region of the jet (solid line) and the background for the same region (dashed line). (Movie 1) \label{fig1}}
\end{figure*}

\clearpage
\begin{figure*}
\epsscale{2}
\plotone{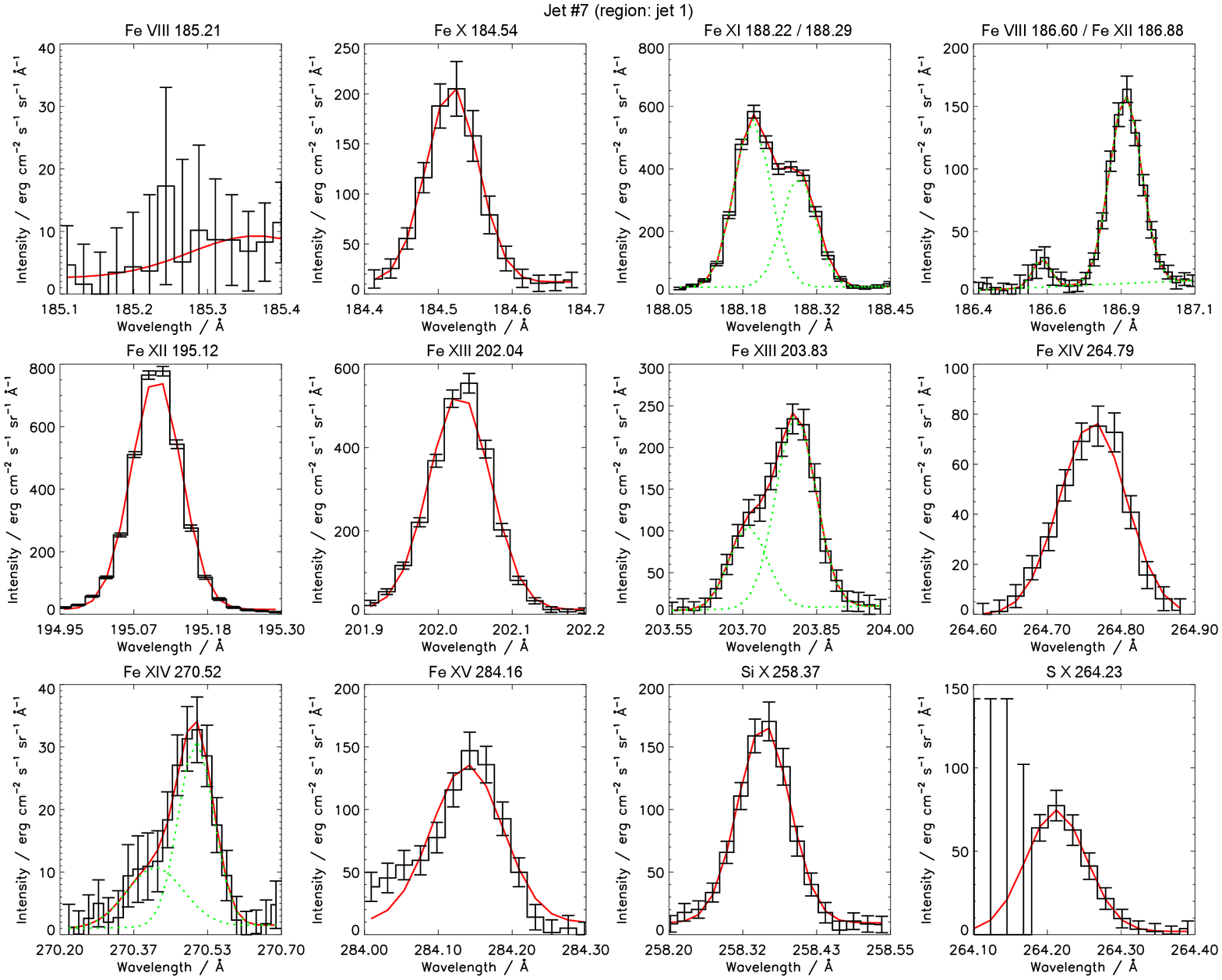} \caption{Spectra derived by subtracting the background spectrum from the jet spectrum for the 'jet 1' region of the jet (\#7 in Table 2) in the upper panel of Figure 1. The red solid lines and green dotted lines represent the total line profiles from single or double Gaussian fitting and each Gaussian component for the double Gaussian fit, respectively. \label{fig2-1}}
\end{figure*}

\clearpage
\begin{figure*}
\epsscale{2}
\plotone{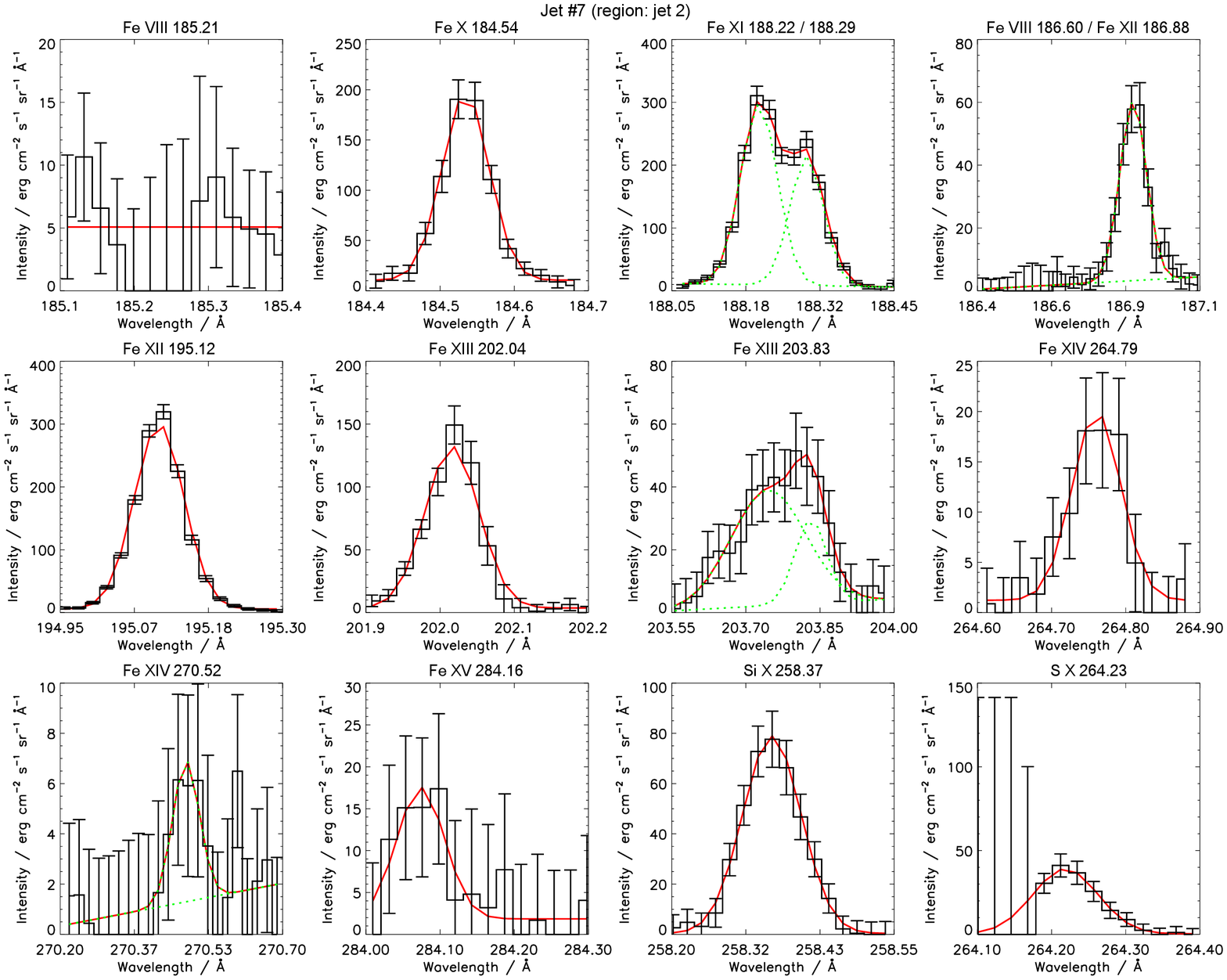} \caption{Spectra derived by subtracting the background spectrum from the jet spectrum for the 'jet 2' region of the jet (\#7 in Table 2) in the upper panel of Figure 1. The red solid lines and green dotted lines represent the total line profiles from single or double Gaussian fitting and each Gaussian component for the double Gaussian fit, respectively. \label{fig3}}
\end{figure*}

\clearpage
\begin{figure*}
\epsscale{2}
\plotone{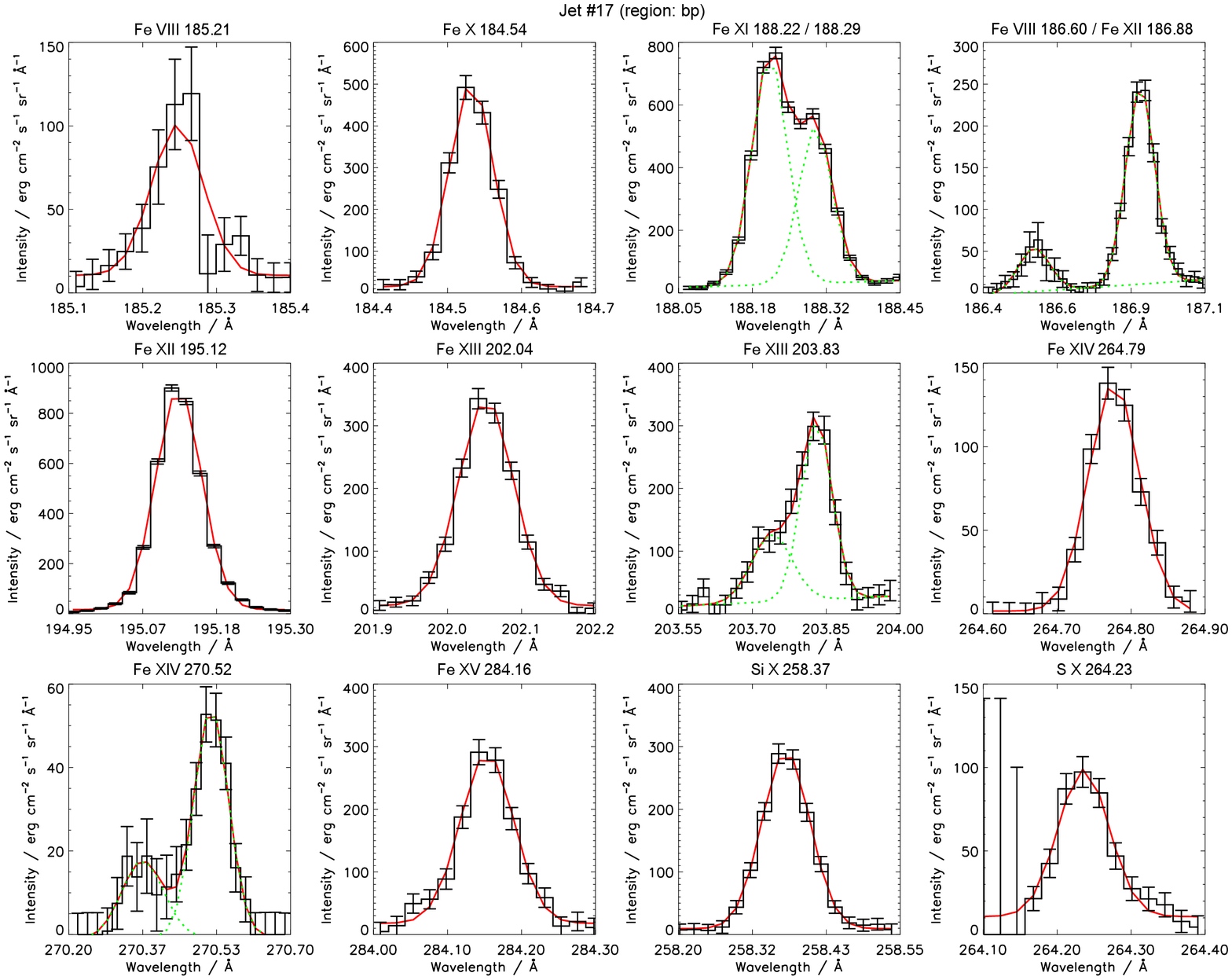} \caption{Spectra derived by subtracting the background spectrum from the jet spectrum for the 'bp' region of the jet (\#17 in Table 2) in the bottom panel of Figure 1. The red solid lines and green dotted lines represent the total line profiles from single or double Gaussian fitting and each Gaussian component for the double Gaussian fit, respectively. \label{fig4}}
\end{figure*}

\clearpage
\begin{figure*}
\centering
\epsscale{2}
\plottwo{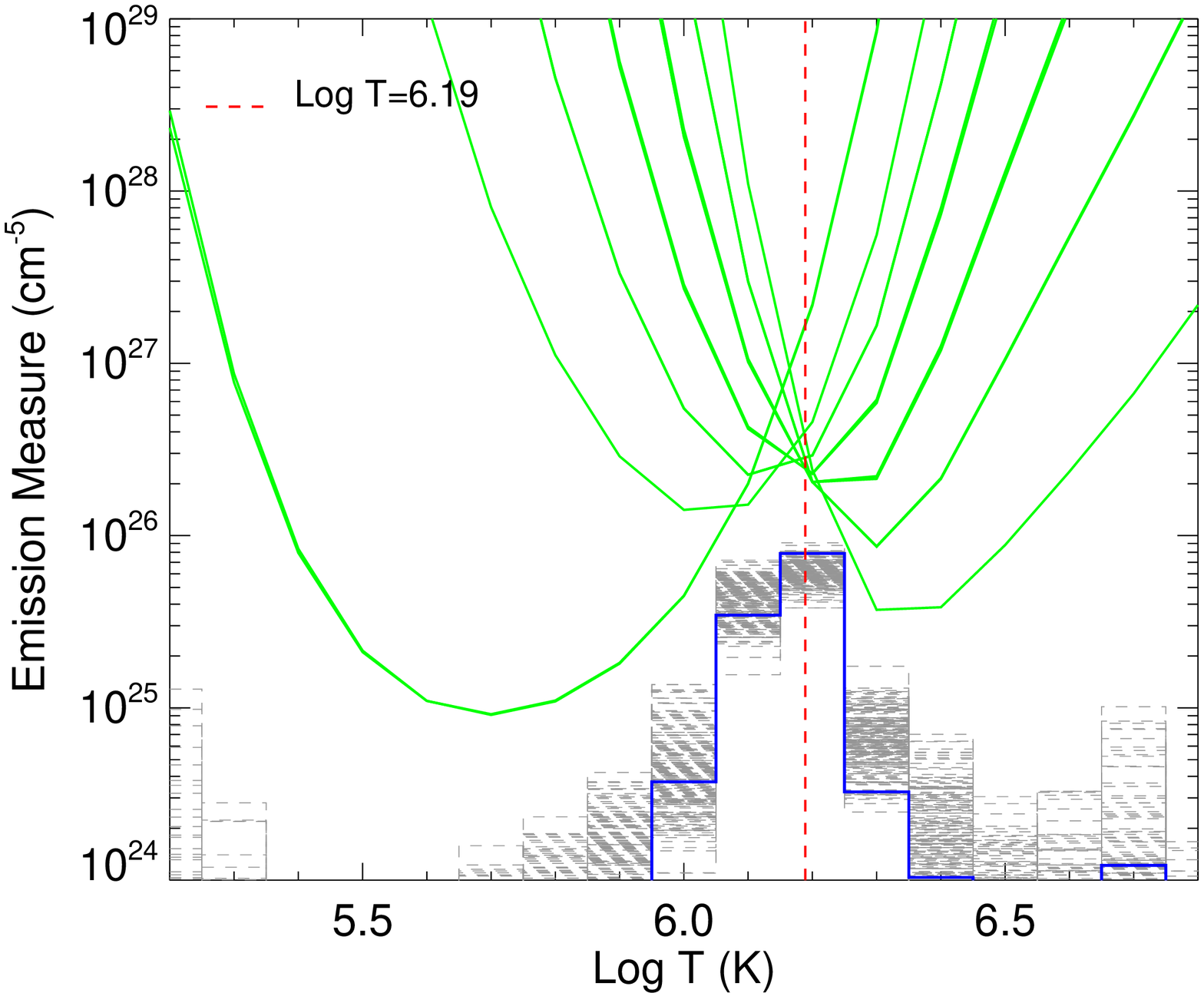}{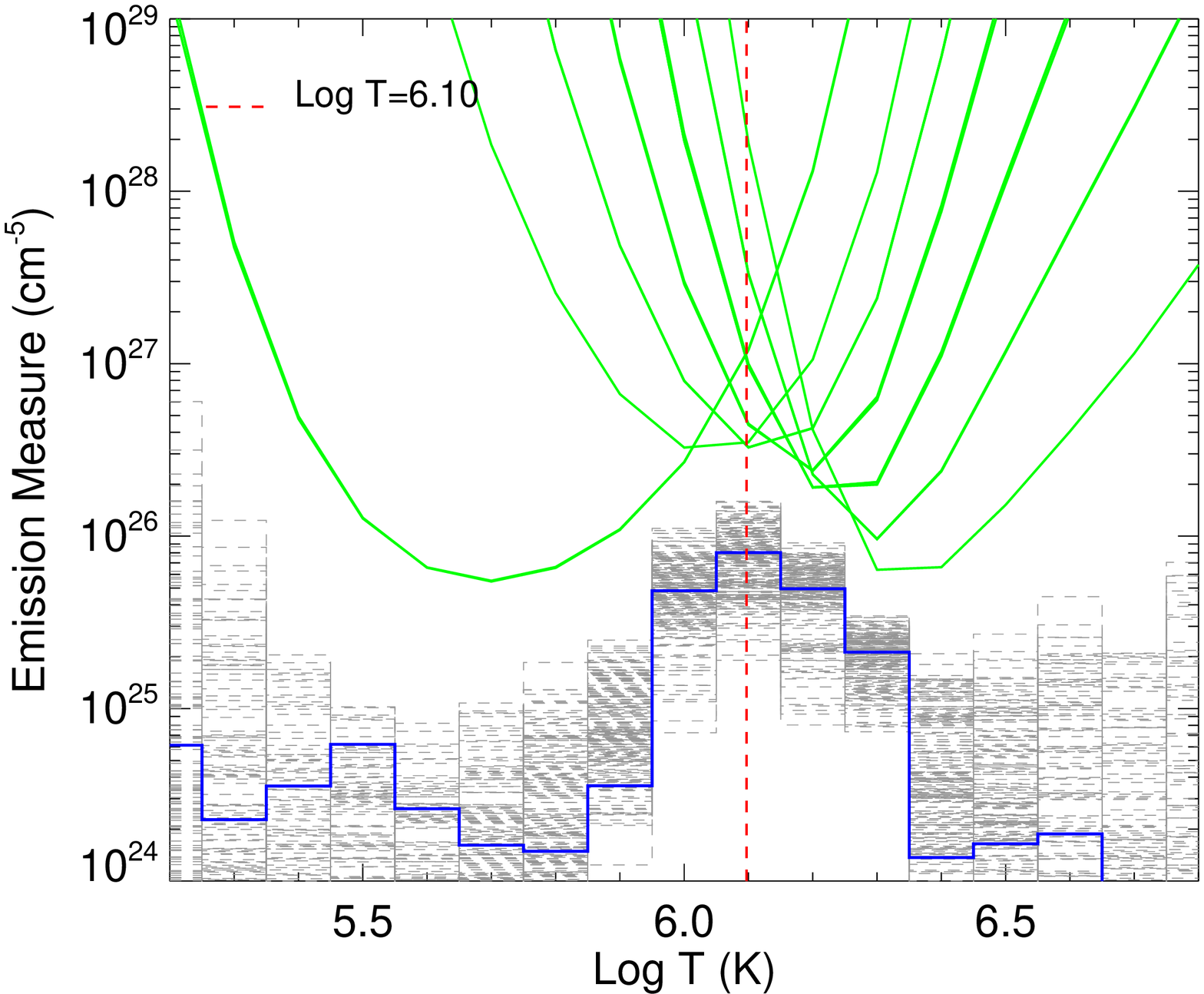} \caption{Emission Measure (EM) distributions from the DEM determined using the background-subtracted intensities of the `jet 1' and `bp' regions of the jets (left: \#7, right: \#17 in Table 2). The grey dashed lines show the results from 200 Monte-Carlo simulations using intensities perturbed within the 1-sigma error, and the blue line represents the best-fit solution. The green lines are EM loci curves and the red dashed line indicates the peak EM temperature. \label{fig5}}
\end{figure*}

\clearpage
\begin{figure*}
\centering
\epsscale{2}
\plotone{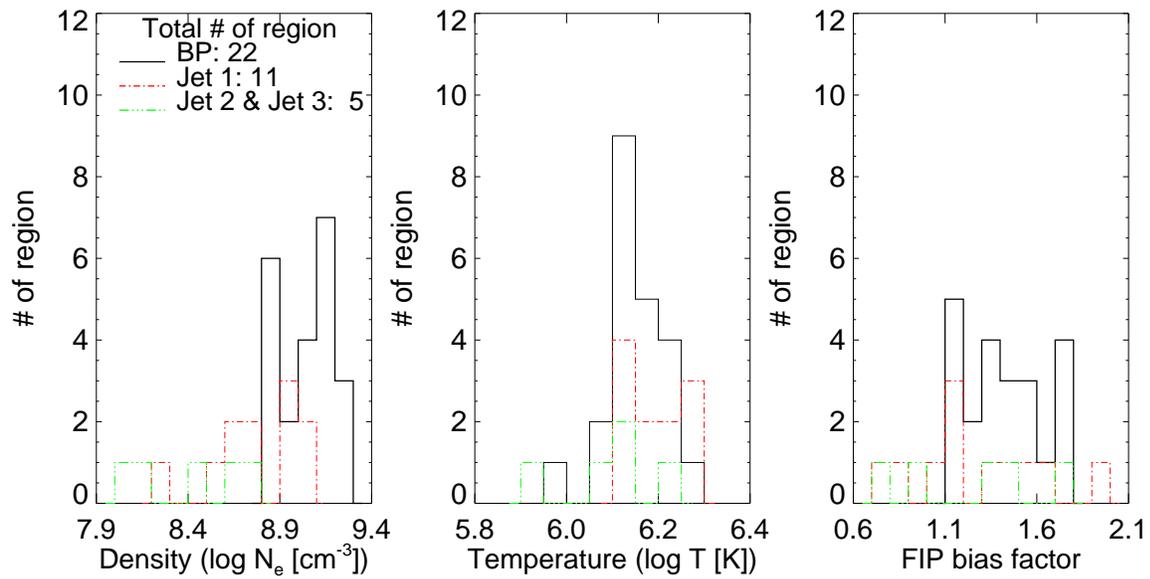} \caption{Histograms of the density, temperature, and FIP bias factors of the polar jets. Black, red, and green lines represent the regions of the polar jets; `bp', 'jet 1', and `jet 2 and jet 3',  respectively. \label{fig6}}
\end{figure*}

\end{document}